\documentclass{article}
\usepackage{spconf,amsmath,graphicx}
\usepackage{cite}
\usepackage{multirow}
\usepackage{array}
\usepackage{makecell}
\usepackage{flushend}
\usepackage{balance}
\usepackage{url}
\usepackage{booktabs}
\usepackage{caption}
\usepackage[svgnames,dvipsnames]{xcolor}

\title{Spoof detection using time-delay shallow neural network and feature switching}
\name{Mari Ganesh Kumar$^1$, Suvidha Rupesh Kumar$^2$, Saranya M S$^1$, B. Bharathi$^2$, Hema A. Murthy$^1$\thanks{We would like to thank the ASV-Spoof-2019 organizers for providing the new spoof detection dataset.}}
\address{ $^1$Indian Institute of Technology Madras\\
  $^2$SSN college of Engineering}
\begin{document}

\maketitle
\begin{abstract}
Detecting spoofed utterances is a fundamental problem in voice-based biometrics. Spoofing can be performed either by logical accesses like speech synthesis, voice conversion or by physical accesses such as replaying the pre-recorded utterance. Inspired by the state-of-the-art \emph{x}-vector based speaker verification approach, this paper proposes a time-delay shallow neural network (TD-SNN) for spoof detection for both logical and physical access. The novelty of the proposed TD-SNN system vis-a-vis conventional DNN systems is that it can handle variable length utterances during testing. Performance of the proposed TD-SNN systems and the baseline Gaussian mixture models (GMMs) is analyzed on the ASV-spoof-2019 dataset. The performance of the systems is measured in terms of the minimum normalized tandem detection cost function (min-t-DCF). When studied with individual features,  the TD-SNN system consistently outperforms the GMM system for physical access. For logical access, GMM surpasses TD-SNN systems for certain individual features. When combined with the decision-level feature switching (DLFS) paradigm, the best TD-SNN system outperforms the best baseline GMM system on evaluation data with a relative improvement of 48.03\% and 49.47\% for both logical and physical access, respectively. 
\end{abstract}

\section{Introduction}
Although automatic speaker verification (ASV) systems are robust to impostor threats~\cite{z-effect-impostor} and acoustic variations,  they are vulnerable when subjected to presentation attacks.  Presenting a fake biometric sample to a biometric detection system is referred to as a  presentation attack\footnote{https://www.iso.org/standard/53227.html}. The process of this deliberate evasion is called spoofing. Spoofing at sample acquisition stage can be classified into two categories, namely, logical access (LA) and physical access (PA)~\cite{zhizhengWu_2015}. Spoofing samples generated using speech synthesis (SS) or voice conversion (VC) approach are categorized as LA while replaying a pre-recorded original audio falls under the PA category. The primary objective of ASV-spoof-challenge proposed in 2015 was to detect LA.  Since the implementation of PA is easier than LA, the former attack is a greater threat than later. ASV-spoof-challenge in 2017 focused on identifying physical access. Numerous spoof detection algorithms have been proposed since then for both LA~\cite{tanvina_asvSpoof_2015,asvSpoof_2015_review,tanvina_asvSpoof_2015_rank1} and PA~\cite{asvSpoof_2017_consolidation,zeroEER_replayAttack,IS2018_dlfs_replayAttacks}. 

ASV-spoof-2019 challenge focused on detecting spoofed utterances synthesized by both LA and PA. Unlike the previous anti-spoofing challenges,  equal error rate (EER) was not used as the evaluation metric due to its ill-suited operating point for user applications like telephone banking~\cite{t-DCF_odyssey}. Hence a new metric termed as a minimum normalized tandem detection cost function (min-t-DCF) is provided as the evaluation metric. The min-t-DCF considers the false alarms and misses for both countermeasure system as well as the automatic speaker verification (ASV) system, along with the prior probabilities of target and spoof trials. The details of min-t-DCF is discussed in ~\cite{tDCF_journal,t-DCF_odyssey}. Scores from a \emph{x}-vector based speaker verification system \cite{Snyder2018XVectorsRD} are used along with the statistics of the spoof detection system to estimate min-t-DCF. 
\emph{x}-vector is a DNN based state-of-the-art speaker verification technique that embeds the speaker characteristics in low-dimensional fixed-length vectors from variable length utterances. 

In this paper, inspired by the \emph{x}-vector based ASV system, we propose a similar spoof detection system for identifying both logical and physical access. For detecting spoofed utterances, the following changes are made to the time-delay neural network architecture (\emph{x}-vector) proposed in~\cite{Snyder2018XVectorsRD}: (i) The last layer in the ASV system's architecture is modified to handle the two-class problem of spoof detection. (ii) Instead of the standard cross-entropy loss function,  a new focal loss function~\cite{focalLossFunction} is used to give more focus on hard and misclassified examples (iii) The network was made shallow since this is binary classification problem with limited data. The proposed network outperforms the baseline GMM classifier for physical access almost in all the cases. On the other hand, the proposed system is not consistently outperforming the GMM classifier while detecting logical access. Instead of conventional score fusion, decision-level feature switching (DLFS) system proposed for ASV-spoof-2017 dataset~\cite{IS2018_dlfs_replayAttacks}  is used to exploit the property of different features in capturing different kinds of spoofing conditions.
The focus of this paper is three-fold: Firstly, a comparison of baseline GMM system using four different features on ASV-spoof-2019 challenge is discussed. Secondly, we propose a novel neural network architecture  for spoof detection system inspired by state-of-the-art ASV system (\emph{x}-vector~\cite{Snyder2018XVectorsRD}). Finally, by using DLFS on individual feature system, the performance of spoof detection systems (SDS) is further improved. 

The rest of the paper is organized as follows: Section~\ref{sec:literatureSurvey} discusses the details of spoof detection approaches in the literature. A brief description of ASV-spoof-2019 dataset is given in Section~\ref{sec:dBdescription}. Section~\ref{sec:asv_spkrRec_xvec} gives a brief overview of the \emph{x}-vector based ASV system. The proposed architecture for spoof detection is explained in Section~\ref{sec:spoofDetection_xvec}. Section~\ref{sec:spoofDetectionSystem} discusses the details of baseline GMM systems, the proposed systems, and the DLFS systems. A comprehensive analysis on the performance of various systems is given in Section~\ref{sec:ResultAnalysis} followed by the conclusion in Section~\ref{sec:conclusion}. 

\section{Prior works on spoof detection}
\label{sec:literatureSurvey}
The ASV-spoof-2015 challenge targeted ten different types of logical access~\cite{asvSpoof_2015}. A combination of auditory transformation based on cochlear filter cepstral coefficients (CFCC) and instantaneous frequency (IF) termed as CFCCIF is proposed as the best feature to detect these LAs in~\cite{tanvina_asvSpoof_2015_rank1}. Score fusion of CFCCIF and MFCC was adjudged as the best system with an average EER of $1.2\%$ across all the ten conditions. Various LA spoof detection systems submitted to the challenge are detailed in~\cite{asvSpoof_2015_review}. 

The speech corpus used in ASV-spoof-2017 challenge has the spoofed instances generated by recording and replaying the bonafide trials of  speakers in different environments (E) using various recording (R) and playback devices (P). Physical attack is harder than logical access as the spoofed utterance of a bonafide trial may come from various E-R-P combinations. The evaluation subset of the ASV-spoof-2017 dataset tried to simulate this `in-wild' condition by generating the spoofed instances from different E-R-P combinations.
A light convolutional neural network (LCNN)~\cite{asvSpoof_rank1} system outperformed all other systems submitted to the challenge. In~\cite{zeroEER_replayAttack} an end-to-end neural network (NN) with attention masking was proposed to learn the difference in the spectrogram of bonafide and the replayed utterances. This end-to-end attention masking system pre-trained on ImageNet dataset~\cite{ImageNet_dataset} gives an ideal performance with zero percent EER. DLFS paradigm proposed in~\cite{IS2018_dlfs_replayAttacks}, uses information from multiple feature spaces. This technique outperforms all other replay attack detection systems in the literature except the ideal NN system with zero percent EER. 

Many recent works on ASV-spoof-2019 dataset uses various end-to-end neural network (NN) structures like DNN with nine layers~\cite{9layer_DNN}, variations of ResNet~\cite{variations_ResNet}, namely, Squeeze-network (SENet), dilated ResNet, and light convolution neural network (LCNN)~\cite{galina_LCNN_asvSpoof2019}. The NN architectures used in these works are deep NNs with a minimum of seven layers excluding the input, pooling, and output layers. The SENet architecture in~\cite{variations_ResNet} uses four blocks of NN architecture with several layers of CNN/RNN in each block. 

\section{Dataset Description}
\label{sec:dBdescription}

Similar to the ASV-spoof-2015 and ASV-spoof-2017 corpus,~\cite{asvSpoof_2017} ASV-spoof-2019 also has three subsets namely, training (train), development (dev), and evaluation (eval). Different subsets of data are used for LA and PA attacks.
The duration of each utterance is approximately two seconds. Unlike the ``in-wild" spoofed trials of the ASV-spoof-2017 corpus, in this dataset, the spoofed trials for physical access are generated in controlled acoustic conditions~\cite{asvSpoof_2019}. The latest best performing text-to-speech synthesis and voice conversion algorithms are used to generate the spoofed trials for logical access category. These algorithms are better than the algorithms used in ASV-spoof-2015. 
The number of trials in each subset is listed in the Table~\ref{tab:datasetDescription}.  The number of trials in evaluation subsets of LA and PA are 71,747 and 137,457, respectively. 

\begin{table}[h]
    \centering
    \caption{Number of trials in development and training subsets}
    \setlength\extrarowheight{2pt}
    \resizebox{0.8\linewidth}{!}{
    \begin{tabular}{c|c|c|c|c|c}
    \hline \hline
        \multirow{2}{*}{Attack} & \multirow{2}{*}{Subsets} & \multicolumn{2}{c}{No. of speakers} & \multicolumn{2}{c}{No. of trials} \\ \cline{3-6}
         & & Male & Female & Bonafide & Spoofed \\ \hline
        \multirow{3}{*}{LA} & train & 8 & 12 & 2580 & 22800 \\ 
             & dev & 8 & 12 & 2548 & 22296 \\ 
             & eval & - & - & 7355 & 63822 \\ \hline
        \multirow{3}{*}{PA} & train & 8 & 12 & 5400 & 48600 \\ 
             & dev & 8 & 12 & 5400 & 24300 \\ 
             & eval & - & - & 18090 & 116640 \\ \hline
    \end{tabular}
    }
    \label{tab:datasetDescription}
\end{table}

\section{\emph{x}-vectors in speaker recognition}
\label{sec:asv_spkrRec_xvec}
{i}-vectors were the state-of-the-approach for text-independent speaker recognition since 2010~\cite{ivec_asv}. An alternate approach proposed in~\cite{snyder2017_IS} extracts DNN embeddings termed as \emph{x}-vectors from a NN using a temporal pooling layer. This pooling layer facilitates the NN to discriminate the speakers from variable-length input speech segments. During testing, the fixed dimensional \emph{x}-vectors are extracted and are compared with the training data embeddings using some scoring approach. 

Speaker embeddings are extracted in ~\cite{snyder2017_IS} from variable length acoustic segments using a DNN with a multi-class cross-entropy loss function. The DNN consists of few time-delay neural network (TDNN) layers to enhance frame-level representation. A pooling layer aggregates the frame-level representations, followed by few additional layers to handle segment-level representations. Finally, a softmax layer is used to get posterior probabilities of each speaker. This approach mainly aims (i) to produce the speaker embeddings at utterance level rather than frame level and (ii) to generalize well, to handle the unseen speakers. The main advantage of this \emph{x}-vector architecture is to handle the short duration utterances. \emph{x}-vector results in~\cite{snyder2017_IS} are shown to outperform the {i}-vector systems for short utterances of duration less than 10 seconds. 

\section{TD-SNN for spoof detection}
\label{sec:spoofDetection_xvec}
Generally, speaker information is present throughout the utterance. Inspired by this concept, x-vector architecture was proposed for automatic speaker recognition in \cite{Snyder2018XVectorsRD,snyder2017_IS}. The x-vector embeddings are obtained by averaging various statistics across time in a high-dimensional space. Similar to the speaker characteristics, the impact of various spoofing approaches used to generate the spoofed trials will be present throughout the utterance. X-vector proposed for ASV \cite{Snyder2018XVectorsRD,snyder2017_IS}, is trained using the speaker labels to extract the speaker embeddings from the data. In this work, we show that the same x-vector model can be used for spoof detection by training the NN using the class labels (bonafide and spoofed) rather than the speaker labels. The results show that the model in fact captures the characteristics of spoofing approaches embedded in the spoofed utterances. 

\emph{x}-vector proposed for ASV in \cite{snyder2017_IS} uses eight hidden layers. Unlike ASV \emph{x}-vector architecture and few other neural network architectures for spoof detection~\cite{galina_LCNN_asvSpoof2019, variations_ResNet, end-to-end-PA_asvSpoof2019, ensembleModels_asvSpoof2019}, we propose a time-delay shallow neural network with just four hidden layers, which includes two hidden layers at frame-level, a pooling layer to aggregate the statistics at the utterance level, and a penultimate layer to reduce the dimension. Time-delay neural network is used for the first time to detect the spoofing attacks.

\begin{figure}
    \centering
    \includegraphics[scale=0.28]{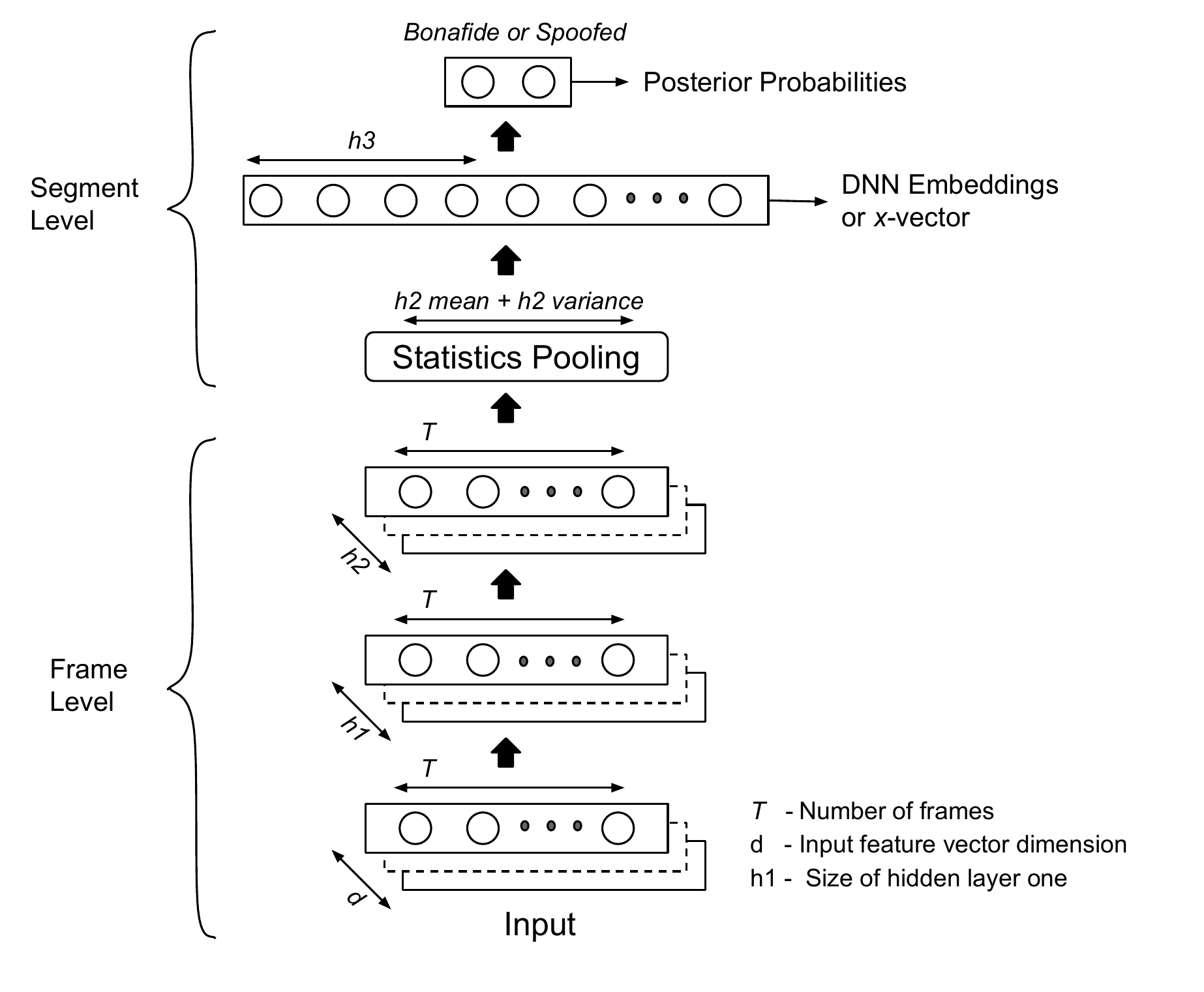}
    \caption{TD-SNN architecture for spoof detection}
    \label{fig:x-arch}
\end{figure}

\begin{table*}[t]
	\centering
	\begin{tabular}{ccc}
		\begin{minipage}{0.6\textwidth}
			\centering
			{\includegraphics[scale=0.37]{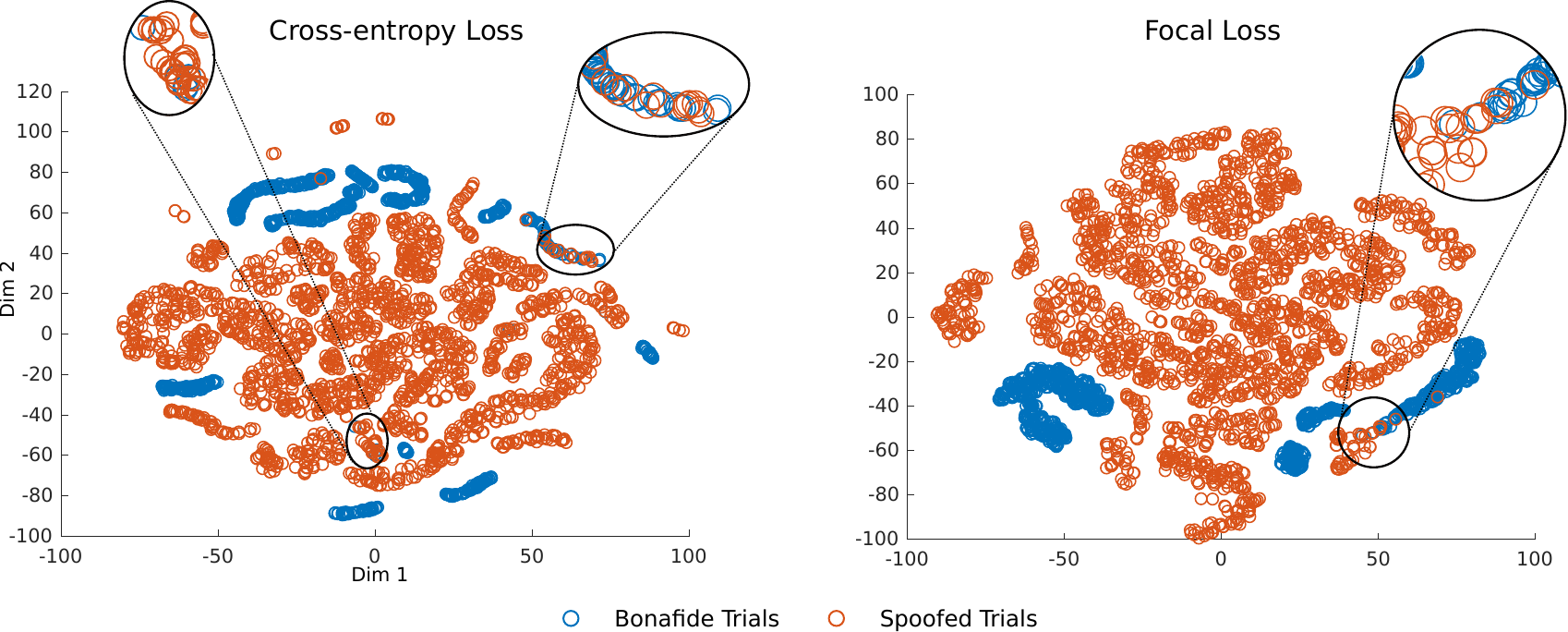}}
			\captionof{figure}{Comparison of proposed network embeddings trained using cross-entropy loss and focal loss. The LA development subset of ASV-spoof-2019 dataset is used to generate this plot. }
			\label{fig:focal}
		\end{minipage} & 
		\begin{minipage}{0.01\textwidth}
			\hfill
		\end{minipage}
		\begin{minipage}{0.35\textwidth}
			\centering
			\vspace{-0.9cm}
			\captionof{table}{List of developed systems.}
	        \resizebox{0.95\textwidth}{!}{
	        \setlength\extrarowheight{3pt}	        
    		    \begin{tabular}{c|c|c|c}
        		    \hline
            		\multirow{2}{*}{Type} & \multirow{2}{*}{System} &  \multicolumn{2}{c}{System Name} \\ \cline{3-4}
	            & & GMM & TD-SNN \\ \hline
        		    \multirow{4}{*}{Single} & \multirow{4}{*}{Baseline} & G-CQCC & x-CQCC  \\
    		        	& &  G-LFCC & x-LFCC \\
	            & & \colorbox{lightgray}{G-IMFCC} & x-IMFCC \\
        		    & & \colorbox{lightgray}{G-LFBE} & x-LFBE \\  \hline
            		\multirow{3}{*}{\makecell{DLFS}} & Primary & \colorbox{lightgray}{ G-Prim} & x-Prim \\ \cline{2-4}
		        & Contrastive-1 & \colorbox{lightgray}{G-C1} & x-C1 \\  \cline{2-4}
        		    & Contrastive-2  & \colorbox{lightgray}{G-C2} & x-C2 \\ \cline{2-4}
        		    & DLFS1  & G-DLFS1 & x-DLFS1 \\ \cline{2-4} 
        		    & DLFS2  & G-DLFS2 & x-DLFS2 \\ \hline
        		    \multicolumn{4}{l}{\makecell{\scriptsize{The systems submitted to ASV-spoof-2019 challenge are} \\ \scriptsize{\hspace{-3cm}highlighted in grey color.}}}
	        \end{tabular}
        		}
		\label{tab:systemsDeveloped}
		\end{minipage} \\ 		
	\end{tabular}
\end{table*}

The proposed architecture for spoof detection is shown in Figure~\ref{fig:x-arch}. This architecture is referred to as time-delay shallow neural network (TD-SNN) in rest of this paper.  The first two layers are frame-level layers and use time-delay neural networks. These layers convert the input feature vectors into high-dimensional vectors by preserving temporal information. The third layer averages information across time by estimating mean and standard deviation, thereby converting the inputs of variable length into a fixed length, high-dimensional vector. The fourth hidden layer reduces this high-dimensional vector to a low-dimensional representation. Since spoof detection is a binary classification problem, a soft-max layer with two outputs is used as the last layer, to get the classification posteriors of a trial. These posterior values are used to classify the trials either as bonafide or spoofed.  The embeddings extracted from the penultimate layer can also be used to identify the spoofed utterances using a back-end classifier.


Instead of the standard cross-entropy error, the focal loss function is used in this work. 
Focal loss was first proposed for object detection task in \cite{focalLossFunction}. The focal loss reshapes the cross-entropy loss such that it gives more importance for hard-to-classify and misclassified examples. The focal loss is a better loss function for the class imbalance problem (Refer to the Table~\ref{tab:datasetDescription} for imbalance in the dataset). The focal loss is estimated as shown in Equation~\ref{eq:focalLoss}.
\begin{align}
\scriptstyle
    \mathcal{F}(p,y) = - \alpha \left[y\;(1-p)^{\gamma}\;log(p)\;+\;(1-y)\; (p)^{\gamma}\; log(p) \right]
    \label{eq:focalLoss}
\end{align}

In Equation~\ref{eq:focalLoss}, $y$ is the ground truth class label, and $p$ is the posterior probability given by a neural network. $\alpha$ and $\gamma$ are hyper-parameters in this loss function. Setting $\gamma$ to zero reduces focal loss to the standard cross-entropy loss. In Figure~\ref{fig:focal}, the 2D representation of DNN embeddings obtained from the proposed network trained using the cross-entropy loss and the focal loss are compared. DNN embeddings are converted to 2D space using t-Distributed Stochastic Neighbor Embedding (t-SNE) algorithm~\cite{t_SNE}. It can be observed that focal loss produces better embeddings with lesser inter-class overlap than the standard cross-entropy error.

In ASV, the \emph{x}-vector architecture uses the raw filter bank energies as the input.  Borrowing from the ASV approaches, the same filter bank energies were given as input to the TD-SNN framework.   As the performance was poor, the focus is shifted to use different features for building a better classifier.

\section{Spoof detection systems (SDS)}
\label{sec:spoofDetectionSystem}
Several attempts have been made to train an efficient classifier for spoof detection. The most common classifiers used for the purpose are GMMs and DNNs. Although there are few works with SVMs~\cite{svm_for_spoofing} and i-vectors~\cite{ivectors_for_spoofDetection}, the performance is worse than that of the GMM and DNN classifiers. Hence in this work, we use both GMM and TD-SNN classifiers to detect spoofed trials. GMM-based systems with a set of features were explored, and best performing four systems were submitted to the ASV-spoof-2019 challenge.  The TD-SNN systems were developed post-challenge. The performance of the TD-SNN systems is compared with the submitted GMM-based SDS using both development and evaluation data. 

\subsection{Single feature systems}
\label{ssec:gmmClassifier}
GMM classifier has been the baseline system for all the ASV spoof challenges conducted from 2015 to 2019.
Bonafide and spoofed trials from the training subset are used to train two GMMs, one for the bonafide ($\lambda_{B}$) and other for the spoofed class ($\lambda_{S}$). 
During testing, a trial $t$ is given to $\lambda_{B}$  and $\lambda_{S}$, and the log-likelihood ($\Lambda$) difference is computed as
\begin{equation}
    \mathcal{S}(t) = \Lambda(\lambda_{B}(t)) - \Lambda(\lambda_{S}(t))
\end{equation}
The log-likelihood difference is considered as the final score for the trial $t$. This simple classifier gave an EER of $1.44\%$ and $7.82\%$ on the evaluation data of ASV-spoof-2015~\cite{tanvina_asvSpoof_2015_rank1} and ASV-spoof-2017~\cite{IS2018_dlfs_replayAttacks} respectively. GMM-SDS with a set of cepstral coefficients and filterbank energies were explored for the ASV-spoof-2019 challenge. The GMM systems with constant-Q cepstral coefficients (CQCC)~\cite{cqcc,cqcc_journal}, inverse Mel frequency cepstral coefficients (IMFCC)~\cite{imfcc_spkrID}, linear frequency cepstral coefficients (LFCC)~\cite{lfcc_mfcc_comparison}, and linear filterbank energy (LFBE) gave better performance than few other features like Mel frequency cepstral coefficients (MFCC), inverse Mel filterbank energies (IMFBE), and Mel filterbank energies (MFBE). 
To compare the performance of TD-SNN systems with that of the baseline GMM systems, TD-SNN systems were also developed with the same set of features. 

\subsection{Feature switching systems}
\label{ssec:dlfs_systems}
Almost every spoof detection system uses a score fusion of many single feature based system as the primary system~\cite{asvSpoof_2015_review,asvSpoof_2017_consolidation}. This clearly shows that different features are required to detect different spoofing conditions.  Instead of the conventional score fusion approach, a decision-level feature switching (DLFS) approach proposed in~\cite{IS2018_dlfs_replayAttacks} is used here. For a given trial, DLFS essentially chooses the decision score from a set of individual features, that has maximum discrimination between the bonafide and the spoofed model. In this work, DLFS is implemented with four best performing individual feature based system for both GMM and TD-SNN frameworks. 
The list of systems developed for this work is listed in Table~\ref{tab:systemsDeveloped}. Features used in primary and contrastive DLFS systems vary for logical access and physical access. Table~\ref{tab:gmmSDS_submittedToChallenge} shows the results of the GMM-based SDS systems submitted to the challenge. 

\begin{table}[h]
\centering
    \caption{Performance (in min-t-DCF) of various GMM-based SDS submitted to ASV-spoof-2019 challenge}
	\resizebox{\linewidth}{!}{
	\setlength\extrarowheight{4pt}
	\begin{tabular}{c|c|c|c|c|c}
	\hline
	\makecell{\bf System \\ \bf Type} & \makecell{\bf Attack \\ \bf Type} & \makecell{\bf System \\ \bf Name} & \bf Feature & {\bf Dev Data} & {\bf Eval Data} \\  \hline
	Single & LA & G-LFBE & LFBE & 0.0077 & 0.2059  \\
	       & PA & G-IMFCC & IMFCC & 0.1396 & 0.1518 \\ \hline
	Primary & LA & G-Prim & CQCC$\mid$LFBE & 0.0002 & 0.1333  \\
	        & PA & G-Prim & \makecell{CQCC$\mid$IMFCC\\ $\mid$LFBE} & 0.1236 & 0.1330 \\  \hline
	Contrastive-1 & LA & G-C1 & IMFCC$\mid$LFBE & 0.0003 & 0.1565 \\
	              & PA & G-C1 & CQCC$\mid$LFCC & 0.1226 & 0.1401 \\  \hline
	Contrastive-2 & LA & G-C2 & CQCC$\mid$LFCC & 0.0013 & 0.2139 \\ 
	              & PA & G-C2 & CQCC$\mid$LFBE & 0.1821 & 0.1672  \\ \hline
\multicolumn{6}{l}{\footnotesize{The symbol `$\mid$` represents exclusive-OR. A$\mid$B implies that either feature A (OR) B will be chosen for each trial.}} \\	              
	\end{tabular}}
	\label{tab:gmmSDS_submittedToChallenge}
\end{table}

\begin{table*}[h]
\centering
    \caption{Performance of various spoof detection systems. Systems with best performance in each category are highlighted in grey color.}
	\resizebox{\textwidth}{!}{
	\setlength\extrarowheight{3pt}
	\begin{tabular}{c|c|c|c|c|c|c||c|c|c|c|c}
	\hline
	\multirow{3}{*}{\makecell{\bf System \\ \bf Type}} & \multirow{3}{*}{\makecell{\bf System \\ \bf Name}} & 	\multicolumn{5}{c||}{\bf Logical access} & \multicolumn{5}{c}{\bf Physical access} \\ \cline{3-12}
	& & \multirow{2}{*}{\bf Feature} & \multicolumn{2}{c|}{\bf Development Data} & \multicolumn{2}{c||}{\bf Evaluation Data} & \multicolumn{2}{c|}{\bf Development Data} & \multicolumn{2}{c|}{\bf Evaluation Data} & \multirow{2}{*}{\makecell{\bf Feature}}\\ \cline{4-11}
	& & & min-t-DCF & EER & min-t-DCF & EER & min-t-DCF & EER & min-t-DCF & EER & \\ \hline
	\multirow{8}{*}{\rotatebox[origin=c]{90}{\makecell{Single}}} & G-CQCC & \multirow{2}{*}{CQCC} & \bf 0.0123 & \bf 0.43 & 0.2366 & 9.57 & \bf 0.1953 & \bf 9.87  & \bf 0.2454 & \bf 11.04 & \multirow{2}{*}{CQCC} \\ 
           & x-CQCC & & 0.0164 & 0.54 & \bf 0.154 & \bf 6.93 & 0.3039 & 12.98 &  0.3148 & 11.83 &  \\ \cline{2-12}
		   & G-LFCC & \multirow{2}{*}{LFCC} & 0.0663 & 2.71 & 0.2116 & 8.09 &0.2555 & 11.96 & 0.3017 & 13.54 & \multirow{2}{*}{LFCC}\\ 
           & x-LFCC & & \bf 0.0062 & \bf 0.28 & \bf 0.164 & \bf 6.29 & \bf 0.1231 & \bf 4.53 & \bf 0.1314 & \bf 4.79 &  \\ \cline{2-12}         
		   & G-IMFCC\textcolor{OliveGreen}{$^\dagger$} & \multirow{2}{*}{IMFCC} & \bf 0.0012 & \bf 0.04 & \bf 0.2401 & \bf 10.62 & 0.2078 & 9.19 & 0.3085 & 12.10 & \multirow{2}{*}{IMFCC}\\ 
           & x-IMFCC & & 0.0285 & 1.08 & 0.4020 & 18.95 & \bf 0.1396 & \bf 5.28 & \bf 0.1518 & \bf 5.58 & \\ \cline{2-12}           
		   & G-LFBE\textcolor{BrickRed}{$^*$} & \multirow{2}{*}{LFBE} & \bf 0.0077 & \bf 0.32 & \bf 0.2059 & \bf 10.65 & \bf 0.2581 & 11.47 & 0.3708 & 15.79 & \multirow{2}{*}{LFBE} \\ 
           & x-LFBE & &  0.0561 & 1.88 & 0.265 & 11.12 & \bf 0.1818 & \bf 7.39 & \bf 0.1766 & \bf 6.99 & \\ \hline    \hline       
	\multirow{8}{*}{\rotatebox[origin=c]{90}{\makecell{DLFS}}} & {G-Prim\textcolor{BrickRed}{$^*$}\textcolor{OliveGreen}{$^\dagger$}} & \multirow{2}{*}{\makecell{CQCC $\mid$ LFBE}} &  \colorbox{lightgray}{\bf 0.0002} & \bf \colorbox{lightgray}{0.01} & \bf 0.1333 & \bf 6.14 & 0.1888 & 8.17 & 0.2767 & 11.28 & \multirow{2}{*}{\makecell{CQCC $\mid$ IMFCC $\mid$ LFBE}} \\ 
           & x-Prim &  & 0.0139 & 0.47 & 0.175 & 8.52 & \bf 0.1236 & \bf 4.85 & \bf 0.133 & \bf 4.91 & \\ \cline{2-12}      
	 & {G-C1\textcolor{BrickRed}{$^*$}\textcolor{OliveGreen}{$^\dagger$}} & \multirow{2}{*}{\makecell{IMFCC $\mid$ LFBE}} & \bf 0.0003 & \bf 0.04 & \bf 0.1565 & \bf 6.46 & 0.1972 & 7.53 & 0.2309 & 9.33 & \multirow{2}{*}{\makecell{CQCC $\mid$ LFCC}} \\ 
           & x-C1 &  & 0.0040 & 0.16  & 0.296 & 14.52 & \bf 0.1226 & \bf 4.56 & \bf 0.140 & \bf 5.05 &\\ \cline{2-12}             
	 & {G-C2\textcolor{BrickRed}{$^*$}\textcolor{OliveGreen}{$^\dagger$}} & \multirow{2}{*}{\makecell{CQCC $\mid$ LFCC}} & \bf 0.0013 & \bf 0.04 & 0.2139 & 9.04 & 0.2329 & 8.48 & 0.3058 & 11.34 & \multirow{2}{*}{\makecell{CQCC $\mid$ LFBE}} \\ 
           & x-C2 &  & 0.0142 & 0.47  & \colorbox{lightgray}{\bf 0.107} &  \colorbox{lightgray}{\bf 5.75} & \bf 0.1821 & \bf 7.54 & \bf 0.167 & \bf 6.46 & \\
           \cline{2-12}      
	 & {G-DLFS1} & \multirow{2}{*}{\makecell{CQCC$\mid$ IMFCC $\mid$ LFCC}} & \bf 0.0026 & \bf 0.19 & 0.2070 & 8.92 & 0.1548 & 7.61 & 0.2260 & 9.99 & \multirow{2}{*}{\makecell{CQCC $\mid$ IMFCC $\mid$ LFCC}} \\ 
           & x-DLFS1 &  &  0.0033 & \bf 0.14  & \bf 0.142 & \bf 7.50 &  \colorbox{lightgray}{\bf 0.1171} & \colorbox{lightgray}{\bf 4.13}  & \bf 0.130 & \bf 5.61 &\\            \cline{2-12} 
 	 & {G-DLFS2} & \multirow{2}{*}{\makecell{LFCC$\mid$ IMFCC $\mid$ LFBE}} & 0.0035 &  0.15 & \bf 0.1780 & \bf7.92 & 0.3209 & 10.61 & 0.2838 & 13.23 & \multirow{2}{*}{\makecell{LFCC $\mid$ IMFCC $\mid$ LFBE}} \\ 
           & x-DLFS2 &  & \bf 0.0166 & \bf 0.31 & 0.208 & 11.21 &  \bf 0.1230 & \bf 4.43  & \colorbox{lightgray}{\bf 0.124} & \colorbox{lightgray}{\bf 4.42} &\\                     
          \hline
           \multicolumn{12}{l}{\footnotesize{\makecell[l]{Systems marked with \textcolor{BrickRed}{$^*$} and \textcolor{OliveGreen}{$^\dagger$} were submitted to ASV-spoof-2019 challenge under LA and PA conditions respectively. The symbol `$\mid$` represents exclusive-OR. A$\mid$B implies that either feature A (OR) B will be chosen for each trial.}}}                       
	\end{tabular}
	}
    \label{tab:devEvalResults_gmmXvec}
\end{table*}

\section{Result Analysis}
\label{sec:ResultAnalysis}
The TD-SNN for LA and PA spoof detection is trained only on the corresponding training subsets. To avoid the problem of over-fitting, twenty percentage of training data is used as the validation subset. This TD-SNN is used to test trials from development and evaluation subset. 
The performance of all SDS on the development and evaluation data are listed in Table~\ref{tab:devEvalResults_gmmXvec}. The best performing system is chosen based on the min-t-DCF metric~\cite{tDCF_journal,t-DCF_odyssey}. G-CQCC and G-LFCC are the single feature based baseline systems provided along with the challenge dataset. Results reported in Table~\ref{tab:devEvalResults_gmmXvec} shows that the performance of the proposed system over the  baseline GMM systems under LA category is not consistent across various features. On the other hand, the TD-SNN systems consistently give good performance for PA than the GMM systems. One possible reason could be that, unlike LA, the PA category have enough amount of data~(refer Table~\ref{tab:datasetDescription}) to train the neural network. Since TD-SNN SDS performs well for all the cases, we can conclude it as a more suitable classifier for detecting physical access spoofing. The performance of the SDS is further improved by applying DLFS as shown in the Table~\ref{tab:devEvalResults_gmmXvec}. 

\begin{table}
\centering
\caption{Relative improvement of t-DCF: Logical Access and Physical Access (evaluation data)}
\resizebox{\linewidth}{!}{
\setlength\extrarowheight{10pt}
	\begin{tabular}{c|c|cc|c}
		\hline
		\makecell{System \\ Type} & \makecell{Attack \\ Type} & \makecell{System \\ Name} & t-DCF & R.I (in \%) \\ \hline \hline
		\multirow{3}{*}{\makecell{DLFS \\ Systems}} & LA & \makecell{G-C2 vs \\ x-C2} &  \makecell{0.2139 vs\\ 0.1070} & 49.97 \\ \cline{3-5}
		& PA & \makecell{G-DLFS2 vs \\ x-DLFS2} &  \makecell{0.2383 vs\\ 0.1240} & 47.96 \\ \hline
		\multirow{2}{*}{\makecell{Best Baseline \\ System vs Best \\ Proposed System}} & LA & \makecell{G-LFBE vs \\ x-C2} &  \makecell{0.2059 vs\\ 0.1070} & 48.03 \\ \cline{3-5}
		& PA & \makecell{G-CQCC vs \\ x-DLFS2} &  \makecell{0.2454 vs\\ 0.1240} & 49.47 \\ \hline		
	\end{tabular}}
	\label{tab:relativeImprovement}
\end{table}

Apart from the systems submitted to the challenge, DLFS with new feature combinations are reported in the table as G-DLFS and x-DLFS. The comparison of best performing TD-SNN system is compared with the corresponding GMM system with same feature combination and the best performing GMM system in Table~\ref{tab:relativeImprovement}. From the result analysis of both LA and PA, we can conclude that TD-SNN framework can be a potential model to detect all types of spoofing attacks. It also justifies our assumption that TD-SNN better identifies the traces of spoof mechanism in the spoofed utterances than the GMM. Moreover, since \emph{x}-vector is the current state-of-the-art for ASV, a spoof detection system with a similar framework, will help us to make a common NN framework for spoof detection as well as speaker recognition. 

\section{Conclusion}
\label{sec:conclusion}
Spoofed utterances contain traces of approaches used to generate them. The ability of \emph{x}-vector based NN to capture the utterance level information is established in the field of speaker verification. Hence, in this work, an attempt has been made to develop spoof detection systems using a similar TDNN framework. A  time-delay shallow neural network (TD-SNN) with focal-loss function is proposed as the neural network architecture for spoof detection. On ASV-spoof-2019 dataset, the proposed TD-SNN based SDS outperforms all the GMM based SDS in case of PA, whereas GMM based SDS performs well for LA in some of the cases. Further, DLFS paradigm is used to improve the performance of single feature based SDS. The best performing TD-SNN SDS with DLFS outperforms the best performing GMM-DLFS SDS with a relative improvement of 48.03\% and 49.47\% for LA and PA in terms of min-t-DCF, respectively.

\bibliographystyle{IEEEtran}
\bibliography{ref_replayIntro}

\end{document}